\documentclass[sigconf]{acmart}
\usepackage{graphicx}
\usepackage{amsmath}
\usepackage{booktabs}
\usepackage{algorithm}
\usepackage{algorithmic}
\usepackage{amsfonts}
\usepackage{multirow}
\usepackage{makecell}
\usepackage{subfigure}
\usepackage{color}
\usepackage{bm}
\usepackage{epstopdf}
\usepackage{url}
\usepackage{balance}
\usepackage{threeparttable}
 
\copyrightyear{2021}
\acmYear{2021}
\setcopyright{acmcopyright}
\acmConference[WSDM '21] {Proceedings of the Fourteenth ACM International Conference on Web Search and Data Mining}{March 8--12, 2021}{Virtual Event, Israel}
\acmBooktitle{Proceedings of the Fourteenth ACM International Conference on Web Search and Data Mining (WSDM '21), March 8--12, 2021, Virtual Event, Israel}
\acmPrice{15.00}
\acmDOI{10.1145/XXXXXX.XXXXXX}
\acmISBN{978-1-4503-8297-7/21/03}

\begin{document}

\title{Bipartite Graph Embedding via Mutual Information Maximization}

\author{Jiangxia Cao$^*$, Xixun Lin$^*$}
\thanks{$^*$Both authors contributed equally and are listed in alphabetical order.}
\affiliation{
  \institution{Institute of Information Engineering, Chinese Academy of Sciences \& School of Cyber Security, University of Chinese Academy of Sciences}
  \country{\{caojiangxia, linxixun\}@iie.ac.cn}
 }

\author{Shu Guo}
\affiliation{
  \institution{National Computer Network Emergency Response Technical Team/Coordination Center of China}
	\country{guoshu@cert.org.cn}
}

\author{Luchen Liu, Tingwen Liu}
\affiliation{
  \institution{Institute of Information Engineering, Chinese Academy of Sciences \& School of Cyber Security, University of Chinese Academy of Sciences}
  \country{\{liuluchen, liutingwen\}@iie.ac.cn}
}

\author{Bin Wang}
\affiliation{
  \institution{Xiaomi AI Lab, Xiaomi Inc.}
  \country{wangbin11@xiaomi.com}
}

\renewcommand{\shorttitle}{BiGI}
 \def\authors{Jiangxia Cao, Xixun Lin, Shu Guo, Luchen Liu, Tingwen Liu, Bin Wang}
\begin{abstract}
Bipartite graph embedding has recently attracted much attention due to the fact that bipartite graphs are widely used in various application domains. Most previous methods, which adopt random walk-based or reconstruction-based objectives, are typically effective to learn local graph structures. However, the global properties of bipartite graph, including community structures of homogeneous nodes and long-range dependencies of heterogeneous nodes, are not well preserved. In this paper, we propose a bipartite graph embedding called \textbf{BiGI} to capture such global properties by introducing a novel local-global infomax objective. Specifically, BiGI first generates a global representation which is composed of two prototype representations. BiGI then encodes sampled edges as local representations via the proposed subgraph-level attention mechanism. Through maximizing the mutual information between local and global representations, BiGI enables nodes in bipartite graph to be globally relevant. Our model is evaluated on various benchmark datasets for the tasks of top-K recommendation and link prediction. Extensive experiments demonstrate that BiGI achieves consistent and significant improvements over state-of-the-art baselines. Detailed analyses verify the high effectiveness of modeling the global properties of bipartite graph. 
\end{abstract}



 \maketitle
\section{Introduction}
Bipartite graph is a general structure to model the relationship between two node types. It has been widely adopted in many real-world applications, arranging from recommender system~\cite{berg2017graph}, drug discovery~\cite{Yamanishi2010DrugtargetIP} to information retrieval~\cite{zhang2019neural}. For instance, in recommender systems, user and item represent two node types. The interactions between users and items are formed as a bipartite graph, where observed edges record previous purchasing behaviours of users. Furthermore, different from heterogeneous graphs, bipartite graph has its own structural characteristics, e.g., there are no direct links between nodes of the same type.
\par
Learning meaningful node representations for bipartite graphs is a long-standing challenge. Recently, a significant amount of progresses have been made toward the graph embedding paradigm~\cite{Hamilton2017RepresentationLO,Cui2017ASO,cai2018comprehensive}. 
Although they work pretty well in the settings of homogeneous and heterogeneous graphs, most of them are not tailored for modeling bipartite graphs. As a result, they are sub-optimal to learn bipartite graph embedding~\cite{Gao2018BiNEBN,Gao2019LearningVR}. To remedy such a problem, several studies have been specifically proposed for modeling bipartite graphs. They can be roughly divided into two branches: random walk-based and reconstruction-based methods. The former~\cite{zhang2017learning,Gao2018BiNEBN,Gao2019LearningVR} relies on designing the heuristics of random walks to generate different node sequences. Afterwards, they learn node representations via
predicting context nodes within a sliding window~\cite{zhang2018arbitrary}. The reconstruction-based works\cite{berg2017graph,He2017NeuralCF,ying2018graph,NGCF19,FOBE,Zhang2019IGMC} are closely related with collaborative filtering~\cite{sarwar2001item}. They attempt to reconstruct the adjacency matrix by learning different encoders. In particular, some works~\cite{berg2017graph,ying2018graph,NGCF19,Zhang2019IGMC} train graph neural networks (GNNs)~\cite{kipf2016semi,gilmer2017neural,lin2020:epo,wu2020comprehensive} to learn node representations via aggregating features of neighborhood nodes recursively. 
\par
Above methods achieve promising results to some extent, but they mainly focus on learning local graph structures with the assumption that nodes within the sliding window or neighborhoods are closely relevant~\cite{Gao2018BiNEBN,NGCF19,Park2019UnsupervisedAM}. We argue that they lack the capability of better modeling the global properties of bipartite graph including community structures of homogeneous nodes and long-range dependencies of heterogeneous nodes. A concrete example is shown in Figure 1. In the user-movie bipartite graph, the movies ``Lion King'', ``Ice Age'' and ``Toy Story'' can be regarded as belonging to the same group since they have similar genres, but the community structure of these three homogeneous nodes is not well preserved by previous methods, due to the fact that ``Lion King'' is unreachable to  ``Ice Age'' and ``Toy Story''. In addition, because ``Lily'' and ``Ice Age'' are distant from each other, the long-range dependency between these two heterogeneous nodes is also hard to be revealed from the local graph structures of them, even ``Lily'' is likely to be interested with ``Ice Age''.
\begin{figure}
\begin{center}
\includegraphics[width=8.5cm,height=4.5cm]{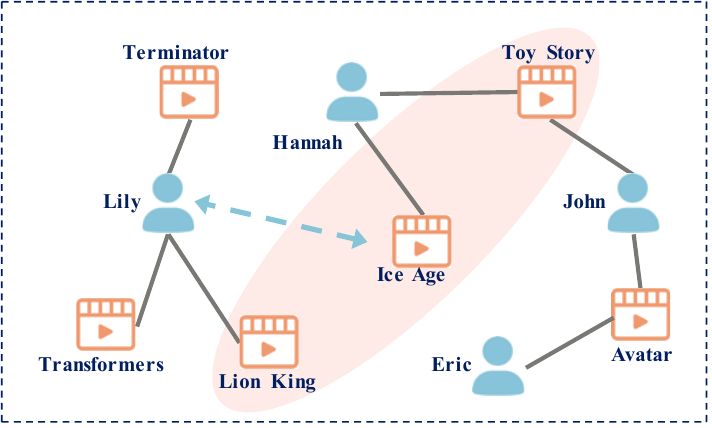}
\caption{An example of user-movie bipartite graph. 
The orange shaded area represents a underlying community structure where three movies may share similar genres. The blue dotted lines denote the long-range dependency between ``Lily'' and ``Ice Age''. However, these global properties are hard to be well learned from local graph structures.}
\label{homo}
\end{center}
\vspace{-3mm}
\end{figure}
\par
To recognize the global properties of bipartite graph, we propose a novel \underline{Bi}partite \underline{G}raph embedding called \textbf{BiGI} via mutual \underline{I}nformation maximization.
Specifically, BiGI first introduces a global representation which is composed of two prototype representations, and each prototype representation is generated by aggregating the corresponding homogeneous nodes. BiGI then encodes sampled edges as local representations via the proposed subgraph-level attention mechanism. On top of that, we develop a novel local-global infomax objective to maximize the mutual information (MI) between local and global representations. In this way, our infomax objective can preserve community structures of homogeneous nodes via maximizing the MI between each node and its homogeneous prototype. Simultaneously, long-range dependencies of heterogeneous nodes are also captured by maximizing the MI between each node and its heterogeneous prototype. The main contributions of our work are as follows,
\begin{itemize}
    \item We propose a novel bipartite graph embedding called BiGI to capture the global properties of bipartite graph including community structures of homogeneous nodes and long-range dependencies of heterogeneous nodes.
    \item A novel local-global infomax objective is developed via integrating the information of two node types into local and global representations. The global representation is composed of two prototype representations, and the local representation is further armed with an h-hop enclosing subgraph to preserve the rich interaction information of sampled edge.
    \item Our model is evaluated on multiple benchmark datasets for the tasks of top-K recommendation and link prediction. Experimental results demonstrate that our method yields consistent and significant improvements over state-of-the-art baselines\footnote{The source code is available from \url{https://github.com/caojiangxia/BiGI}.}.
\end{itemize}
\section{Related Work}
\subsection{Bipartite Graph Embedding}
Homogeneous and heterogeneous graph embeddings are usually used for modeling bipartite graphs. The pioneering homogeneous graph methods include DeepWalk~\cite{Perozzi2014DeepWalkOL}, LINE~\cite{Tang2015LINELI}, Node2vec~\cite{Grover2016node2vecSF} and VGAE~\cite{kipf2016variational}. Some representative heterogeneous graph methods are Metapath2vec~\cite{dong2017metapath2vec} and DMGI~\cite{Park2019UnsupervisedAM}. But they are not tailored for bipartite graphs, and the structural characteristics of bipartite graph are hard to be preserved by them. IGE~\cite{zhang2017learning}, PinSage~\cite{ying2018graph}, BiNE~\cite{Gao2018BiNEBN} and FOBE~\cite{FOBE} are specially designed for bipartite graphs.
However, as mentioned in the \emph{introduction}, they mainly focus on how to model local graph structures in the latent space.
\par
Matrix completion~\cite{berg2017graph,Zhang2019IGMC} and collaborative filtering~\cite{He2017NeuralCF,NGCF19} are also connected with modeling bipartite graphs closely. They propose various DNNs to solve recommendation tasks. 
For example, GC-MC~\cite{berg2017graph} uses one relation-aware graph convolution layer to learn node embeddings, thus only the direct links in user-item bipartite graphs are exploited. NGCF~\cite{NGCF19} incorporates collaborative signals into the embedding process by aggregating features of neighborhood nodes. However, it still overlooks the importance of modeling the global properties of bipartite graph.



\par
\subsection{Mutual Information Maximization}
Maximizing the MI between inputs and corresponding latent embeddings provides a desirable paradigm for the unsupervised learning~\cite{Oord2018RepresentationLW}. However, estimating MI is generally intractable in high-dimensional continuous settings~\cite{Paninski2002EstimationOE}. 
MINE~\cite{Belghazi2018MutualIN} derives a lower bound of MI and works by training a discriminator to distinguish samples coming from the joint distribution of two random variables or the product of their marginals. DIM~\cite{Hjelm2018LearningDR} introduces the structural information into input patches and adopts different infomax objectives.
\par
DGI~\cite{Velickovic2018DeepGI} is the first work that applies the infomax objective to homogeneous graphs. It provides a new approach for the task of unsupervised node classification. 
Based on DIM, InfoGraph~\cite{sun2019infograph} tries to learn unsupervised graph representations via maximizing the MI between the graph-level representation and the representations of substructures. DMGI~\cite{Park2019UnsupervisedAM} extends DGI into heterogeneous graphs. It splits the original graph into multiple homogeneous ones and adopts the infomax objective used in DGI for modeling split graphs. So DMGI still puts more emphasis on learning the correlation of homogeneous nodes. GMI~\cite{peng2020graph} proposes a new approach to measure the MI between input homogeneous graphs and node embeddings directly. Compared with them, we combine two types of node information for generating local and global representations and develop a novel infomax objective that is more suitable for bipartite graphs. 
\begin{figure}
\begin{center}
\includegraphics[width=8.5cm,height=7.2cm]{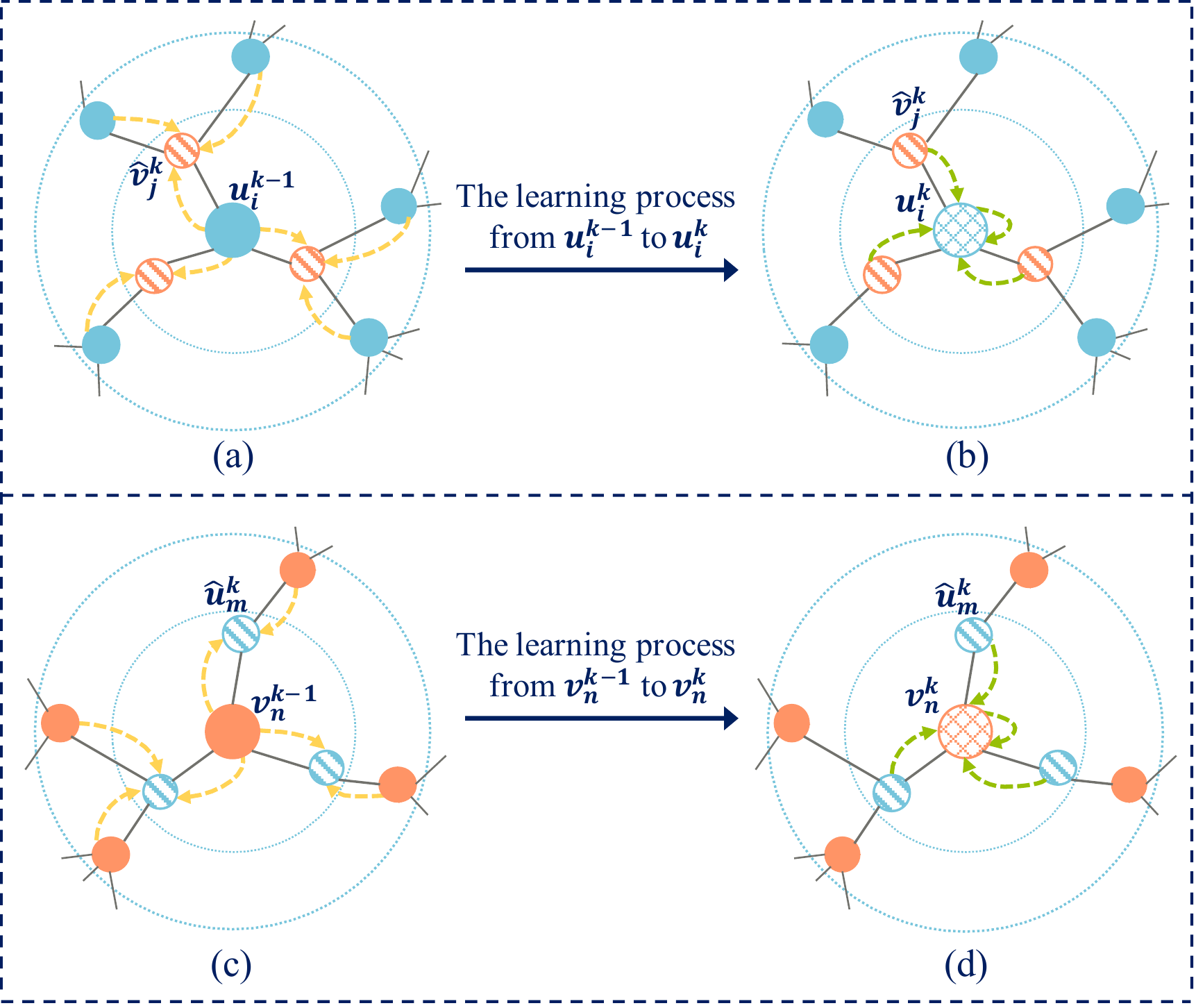}
\caption{A simple illustration of the proposed encoder. In $k$-th layer, (a) and (b) show the learning process of $\bm{u}^{k-1}_i$. (c) and (d) show the learning process of $\bm{v}^{k-1}_n$ in a similar way. The yellow dotted lines (Eq.(\ref{node_type_conversion_1}) and Eq.(\ref{node_type_conversion_2})) and the green dotted lines (Eq.(\ref{convolution_1}) and Eq.(\ref{convolution_2})) demonstrate how to derive node embeddings $\bm{u}^{k}_i$ and $\bm{v}^{k}_n$.}
\label{gnnencoder}
\end{center}
\vspace{-3mm}
\end{figure}
\section{Background}
We begin by providing the background of our work. Let $G=(U, V, E)$ be a bipartite graph, where $U$ and $V$ are two disjoint node sets, and $E \subseteq U \times V$ denotes the edge set. It is obvious that $G$ has two node types. The nodes that fall into the same node set are homogeneous, and the nodes belonging to different node sets are heterogeneous. 
$A \in \{0,1\}^{|U| \times |V|}$ is a binary adjacency matrix, where each element $A_{i,j}$ describes whether node $u_{i} \in U$ has interacted with node $v_{j} \in V$. 
Given a bipartite graph $G=(U, V, E)$ with the adjacency matrix $A$, the goal of bipartite graph embedding is to map each node in $G$ to a $d$-dimensional vector. To keep notations simple, we use $\bm{u}_{i}$ and $\bm{v}_{j}$ to represent the embedding vectors of $u_{i}$ and $v_{j}$, respectively.

\section{Proposed Model}
A novel bipartite graph embedding termed as \textbf{BiGI} is proposed from the perspective of mutual information maximization. We first describe a basic bipartite graph encoder to generate the initial node representations. Taking these node representations as the inputs of our framework, we then demonstrate how to construct the global representation, local representations and the local-global infomax objective. The detailed model analysis is provided in the end.
\subsection{Bipartite Graph Encoder}
In this section, we introduce a basic bipartite graph encoder following the principle of GNN to learn the initial node representations. The proposed encoder is well compatible with our infomax objective. Compared with other GNN encoders~\cite{ying2018graph,NGCF19} for bipartite graphs, it achieves promising performances empirically.

\par
Different from homogeneous graphs, each node in bipartite graph is not the same type as its adjacent nodes. Therefore, directly updating the node embedding via aggregating features of its one-hop neighbors is ill-posed. To alleviate such an issue, our encoder attempts to learn each node embedding from its two-hop neighbors in each layer. As shown in Figure~\ref{gnnencoder}, 
both of the learning processes of $\bm{u}^{k-1}_i$ and $\bm{v}^{k-1}_n$ have two operations in the $k$-th layer. Taking $\bm{u}^{k-1}_i$ for example ((a) and (b) in Figure~\ref{gnnencoder}), we first generate temporary neighborhood representations, e.g., $\widehat{\bm{v}}^{k}_{j}$ via a mean operation (MEAN) with a non-linear transformation:
\begin{equation}
\begin{split}
\widehat{\bm{v}}^{k}_{j} = \delta \Big( \widehat{W}^{k}_{v} \cdot {\rm MEAN} \big( \{ \bm{u}^{k-1}_{i} : u_i \in \mathcal{N}(v_j) \} \big) \Big),
\label{node_type_conversion_1}
\end{split}
\end{equation}
where $\delta$ denotes the ${\rm LeakyReLU}$ activation function, $\widehat{W}^{k}_{v}$ is a weight matrix and $\mathcal{N}(v_j)$ denotes one-hop neighbors of $v_{j}$. In contrast with common graph convolutional operators~\cite{kipf2016semi,hamilton2017inductive,gilmer2017neural}, we only aggregate neighborhood features, and the own feature $\bm{v}^{k-1}_j$ is not involved in Eq.(\ref{node_type_conversion_1}). Hence, $\widehat{\bm{v}}^{k}_{j}$ can be approximately regarded as a $u$-type node embedding. Afterwards, we use homogeneous graph convolution to obtain $\bm{u}^{k}_i$:
\begin{equation}
\begin{split}
\overline{\bm{u}}^{k}_{i} = \delta \Big( \overline{W}^{k}_{u} &\cdot {\rm MEAN} \big( \{ \widehat{\bm{v}}^{k}_{j} : v_j \in \mathcal{N}(u_i) \} \big) \Big),\\
&\bm{u}^{k}_{i} = W^{k}_{u} \cdot \big[\overline{\bm{u}}^{k}_{i} \big| \bm{u}^{k-1}_{i} \big],
\label{convolution_1}
\end{split}
\end{equation}
where $\overline{W}^{k}_{u}$ and $W^{k}_{u}$ are two weight matrices and $[\cdot|\cdot]$ is a concatenation operation. The similar procedures are also employed to update $\bm{v}^{k-1}_n$. Sub-figure (c) illustrates the neighborhood aggregation of $\widehat{\bm{u}}^{k}_{m}$: 
\begin{equation}
\begin{split}
\widehat{\bm{u}}^{k}_{m} = \delta \Big( \widehat{W}^{k}_{u} \cdot {\rm MEAN} \big( \{ \bm{v}^{k-1}_{n} : v_n \in \mathcal{N}(u_m) \} \big) \Big).
\label{node_type_conversion_2}
\end{split}
\end{equation}
The final node embedding $\bm{v}^{k}_{n}$ is defined as: 
\begin{equation}
\begin{split}
\overline{\bm{v}}^{k}_{n} = \delta \Big( \overline{W}^{k}_{v} &\cdot {\rm MEAN} \big( \{ \widehat{\bm{u}}^{k}_{m} : u_m \in \mathcal{N}(v_n) \} \big) \Big),\\
&\bm{v}^{k}_{n} = W^{k}_{v} \cdot \big[\overline{\bm{v}}^{k}_{n} \big| \bm{v}^{k-1}_{n} \big].
\label{convolution_2}
\end{split}
\end{equation}
$\widehat{W}^{k}_{u}$, $\overline{W}^{k}_{v}$ and $W^{k}_{v}$ in Eq.(\ref{node_type_conversion_2}) and Eq.(\ref{convolution_2}) are also weight matrices. Dropout~\cite{srivastava2014dropout} is applied to each layer of our encoder to regularize model parameters. 
\subsection{Local-Global Infomax}
Building upon the generated node representations, in this section, we first present the calculations of global and local representations. A novel local-global infomax objective is then developed to capture the global properties of bipartite graph. 
\subsubsection{Global Representation}
The global representation is a holistic representation of bipartite graph, which is generated via a simple composition function (${\rm COM}$) that combines two prototype representations. Specifically, for each node type, we introduce a prototype representation to aggregate all homogeneous node information. Our insight is similar to the classic few-shot learning~\cite{snell2017prototypical} which would generate a prototype representation of each class. There are many choices to induce the prototype representation. In our work, we also adopt the mean operation which averages the information of all homogeneous nodes to obtain the corresponding prototype representation. The concrete procedures can be formulated as follows,
\begin{equation}
\begin{split}
\bm{p}_{u} = {\rm MEAN} \big(\{\bm{u}_{i} &: u_i \in U\}\big), \quad \bm{p}_{v} = {\rm MEAN}  \big(\{\bm{v}_{i}: v_i \in V\}\big), \\
\bm{g} =~&{\rm COM}\big(\bm{p}_{u}, \bm{p}_{v}\big) = \big[\sigma (\bm{p}_{u}) \big| \sigma (\bm{p}_{v}) \big],
\label{graph_representation}
\end{split}
\end{equation}
where $\bm{u}_{i}$ and $\bm{v}_{i}$ denote the outputs of our encoder. $\bm{g}$ is the global representation composed of two prototype representations $\bm{p}_{u}$ and $\bm{p}_{v}$. For efficiency, we select the simple concatenation operation with the sigmoid activation function $\sigma$ as our composition function. 
\subsubsection{Local Representation}
Each input of local representation is a bipartite edge i.e., $(u,v)$, and we further arm it with an h-hop enclosing subgraph~\cite{Zhang2019IGMC} to describe the surrounding environment of $(u,v)$. The concrete definition of h-hop enclosing subgraph is given below.  
\begin{definition}
\label{def:subgraph}
\textbf{(H-hop Enclosing Subgraph)}
Given a bipartite graph $G=(U, V, E)$, two nodes $u \in U$ and $v \in V$, the h-hop enclosing subgraph for $(u, v)$ is the subgraph $G^{h}_{(u,v)}$ induced from $G$ by the union of two node sets, i.e., $G^{h}(u) \cup G^{h}(v)$. Here, $G^{h}(u) = \{v_i | dis(v_i, u) \leq h \}$, $G^{h}(v) =  \{u_i | dis(u_i, v) \leq h \}$ and $dis$ is a distance function. Due to the particular structure of bipartite graph, $h$ is set as an odd number strictly.
\end{definition}
For a specific edge $(u, v) \in E$ with the corresponding h-hop enclosing subgraph $G^{h}_{(u,v)}$ (The subscripts of $u$ and $v$ are omitted for simplicity), we use an attention mechanism (${\rm ATT}$) to calculate the local representation. 
Given node $u$ and node $v_{i} \in G^{h}(u)$, the attention weight $\alpha_{u,i}$ can be expressed as:
\begin{equation}
\begin{split}
\alpha_{u,i} =  \frac{{\rm exp}\Big\{\big(W_{a} \cdot \bm{v}_{i}\big)^{T} \cdot \big(W_{a}^{\prime} \cdot \bm{u}\big)\Big\}}{\sum_{v_j \in G^{h}(u)}{\rm exp}\Big\{\big(W_{a} \cdot \bm{v}_{j}\big)^{T} \cdot \big(W_{a}^{\prime} \cdot \bm{u}\big)\Big\}},
\label{attention_1}
\end{split}
\end{equation}
where $T$ denotes the transpose operation, $W_{a}$ and $W_{a}^{\prime}$ are two shared trainable matrices. The similar calculation procedure for node $v$ and node $u_{i} \in G^{h}(v)$ can be defined as:
\begin{equation}
\begin{split}
\alpha_{v,i} =  \frac{{\rm exp}\Big\{\big(W_{a}^{\prime} \cdot \bm{u}_{i}\big)^{T} \cdot \big(W_{a} \cdot \bm{v}\big)\Big\}}{\sum_{u_j \in G^{h}(v)}{\rm exp}\Big\{\big(W_{a}^{\prime} \cdot \bm{u}_{j}\big)^{T} \cdot \big(W_{a} \cdot \bm{v}\big)\Big\}}.
\label{attention_2}
\end{split}
\end{equation}
The final representation of local input $\bm{g}_{(u,v)}^{h}$ is formulated as:
\begin{equation}
\begin{split}
\bm{g}_{(u,v)}^{h} = \Big[\sigma\big(\sum_{v_i \in G^{h}(u)} \alpha_{u,i} \bm{v}_{i} + \bm{u}\big) \Big| \sigma \big(\sum_{u_i \in G^{h}(v)} \alpha_{v,i} \bm{u}_{i} + \bm{v}\big) \Big].
\label{subgraph_representation}
\end{split}
\end{equation}
The local attentive representation also combines different local environments together via the same composition function used in Eq.(\ref{graph_representation}). It not only highlights the central role of $(u, v)$, but also adaptively assigns different importance factors to neighboring nodes by the subgraph-level attention mechanism.  
\begin{figure}
\setlength{\abovedisplayskip}{5pt}
\setlength{\belowdisplayskip}{5pt}
\begin{center}
\includegraphics[width=8.2cm,height=7cm]{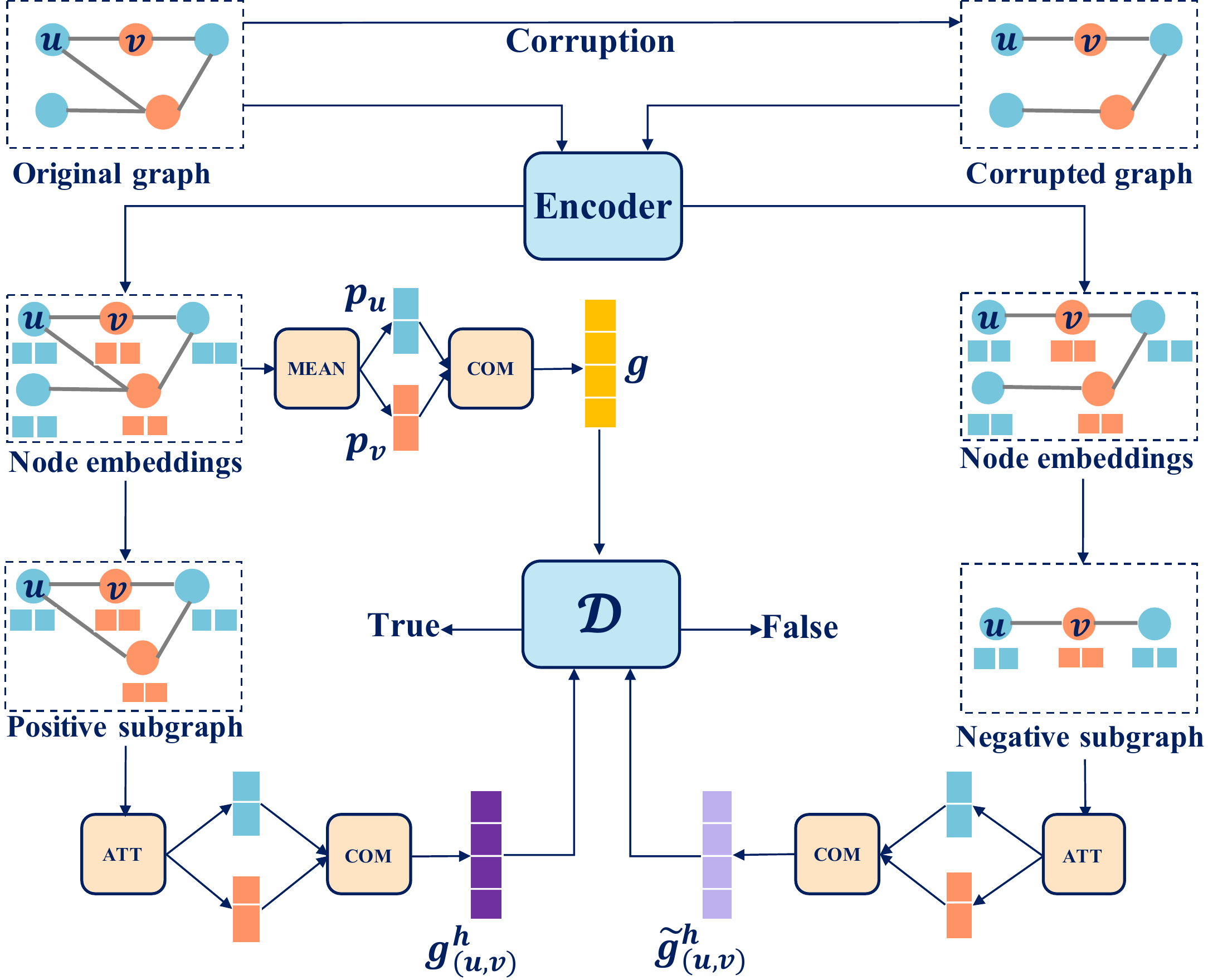}
\caption{An overview of BiGI. ``ATT'', ``MEAN'' and  ``COM'' denote the subgraph-level attention mechanism, mean operation and composition function, respectively. $\bm{p}_{u}$ and $\bm{p}_{v}$ are two prototype representations. $\bm{g}$ is the global representation. $\bm{g}_{(u,v)}^{h}$ and $\bm{\widetilde{g}}_{(u,v)}^{h}$ are local representations.}
\label{BigIM_framework}
\end{center}
\vspace{-3mm}
\end{figure}
\subsubsection{Infomax Objective}
After obtaining local and global representations, 
our local-global infomax objective is reformulated as a noise-contrastive loss, where positive samples come from the joint distribution and negative samples come from the product of marginals. A corruption function $\mathcal{C}$ is required to generate the negative samples, and BiGI uses a general trick that corrupts the graph structure $A$ to define $\mathcal{C}$. The switch parameter $S_{i,j}$ determines whether to corrupt the entry of adjacency matrix $A_{i,j}$. Above operations are performed as follows,
\begin{equation}
\begin{split}
&S_{i,j} = {\rm Bernoulli}(\beta), \\
\widetilde{G} = &(U,V,\widetilde{E}) = \mathcal{C}(G, \beta)= A \oplus S,
\end{split}
\label{corruption}
\end{equation}
where $\beta$ is the corruption rate, $\oplus$ denotes the XOR (exclusive-OR) operation, $\widetilde{G}$ is the corrupted graph and $\widetilde{E}$ is the corresponding set of corrupted edges. The concrete loss function is defined as:
\begin{equation}
\begin{split}
\mathcal{L}_{m} = &-\frac{1}{|E|+|\widetilde{E}|}\Big(\sum_{i=1}^{|E|} \mathbb{E}_{G}\big[{\rm log}\mathcal{D}(\bm{g}_{(u,v)_{i}}^{h}, \bm{g})\big]  + \\ &\sum_{i=1}^{|\widetilde{E}|} \mathbb{E}_{\widetilde{G}}\big[{\rm log}\big(1 - \mathcal{D}(\bm{\widetilde{g}}_{(u,v)_{i}}^{h}, \bm{g})\big) \big] \Big).
\label{MI_graphtosubgraph}
\end{split}
\end{equation}
Here, $\mathcal{D}$ is a discriminator to score local-global representations via a bilinear mapping function:
\begin{equation}
\begin{split}
\mathcal{D}(\bm{g}_{(u,v)_{i}}^{h}, \bm{g}) = \sigma\big((\bm{g}_{(u,v)_{i}}^{h})^{T}W_{b}\bm{g}\big),
\label{bilinear}
\end{split}
\end{equation}
where $W_{b}$ is a weight matrix. The binary cross-entropy loss in Eq.(\ref{MI_graphtosubgraph}) is an effective MI estimator. It can maximize the MI between $\bm{g}^{h}_{(u,v)_{i}}$ and $\bm{g}$, based on Jensen--Shannon divergence between the joint distribution and the product of marginals. Because it follows a standard minmax game originated from the generative adversarial network (GAN)~\cite{goodfellow2014generative}, and the ``GAN'' distance and Jensen-Shannon divergence are closely related~\cite{nowozin2016f}. 
\par
From Eq.(\ref{graph_representation}), we can observe that the information of two node types is integrated into the global representation via the generated prototype representations, and these two prototypes are not entangled together. Through Eq.(\ref{MI_graphtosubgraph}) and Eq.(\ref{bilinear}), each node has access to the homogeneous prototype and to the heterogeneous prototype simultaneously, which enables our model to break the limit of local graph topology. Therefore, the global properties can be naturally captured even the correlated nodes in bipartite graph are distant from each other.  
\subsection{Model Training}
The total loss function $\mathcal{L}$ contains two terms:
\begin{equation}
\begin{split}
\mathcal{L} = \lambda \mathcal{L}_{m} + (1-\lambda) \mathcal{L}_{r} ,
\label{totalloss}
\end{split}
\end{equation}
where $\lambda$ is the harmonic factor. $\mathcal{L}_{r}$ is a margin-based ranking loss over observed edges for our encoder, which is formulated as follows,
\begin{equation}
\begin{split}
\mathcal{L}_{r} = \sum_{(u,v)\in E}\sum_{(u^{\prime}, v^{\prime}) \in E^{\prime}_{(u,v)}}\Big[\gamma + \phi\big([\bm{u^{\prime}} |\bm{v^{\prime}}]\big) - \phi\big([\bm{u} | \bm{v}]\big) \Big]_{+},
\label{gnnloss}
\end{split}
\end{equation}
where $\phi$ is a ranking function parameterized by a two-layer multilayer perceptron (MLP), $[x]_{+}$ denotes the positive part of $x$ and $\gamma$ is a margin. $E^{\prime}_{(u,v)}$ is the set of negative node pairs, which can be defined as:
\begin{equation}
\begin{split}
E^{\prime}_{(u,v)} = \big\{(u^{\prime}, v) |u^{\prime} \in U \big\}\cup\big\{(u, v^{\prime}) |v^{\prime} \in V \big\}.
\label{negativeset}
\end{split}
\end{equation}
The negative sampling used in Eq.(\ref{negativeset}) is similar to~\cite{lin2019:AKE,guo2020:slre}: $E^{\prime}_{(u,v)}$ is composed of real interactions with either the head or tail replaced by a random node from the same node set. BiGI is an end-to-end model which is optimized by Adam~\cite{Kingma2014AdamAM}. The whole architecture of BiGI is shown in Figure~\ref{BigIM_framework}. 
\subsection{Model Analysis}
\subsubsection{Time Complexity}
The main operations of BiGI are learning the initial node representations and calculating the total loss. To avoid parameter overhead, we use the shared encoder to learn node representations of $G$ and $\widetilde{G}$. The computational complexity of BiGI is approximated as $O( k(|E|+|\widetilde{E}|)d^2)$, where $k$ is the number of layers in our encoder and $d$ is the embedding size. In addition, we also provide an experimental comparison to verify that our model can be deployed to large-scale bipartite graphs in Section 5.3. 
\subsubsection{Relation with DGI}
Our model is closely related to DGI, since they use a local-global infomax objective on graphs. However, there are important design differences between them. 1) BiGI focuses on modeling bipartite graphs, which integrates the information of two node types into local and global representations. By contrast, DGI is designed for homogeneous node embeddings. 2) DGI tries to maximize the MI between node-level and graph-level representations, while we actually maximize the MI between subgraph-level and graph-level representations. The subgraph-level representations are capable of effectively preserving rich interactions of sampled edges. 3) The choice of encoders is different. Considering the structural characteristics of bipartite graphs, we design a novel basic encoder to learn initial node representations.
\section{Experiments}

\subsection{Datasets}
Four benchmark datasets, i.e., DBLP~\footnote{https://github.com/clhchtcjj/BiNE/tree/master/data/dblp}, MovieLens-100K (ML-100K)~\footnote{https://grouplens.org/datasets/movielens/100k/}, MovieLens-10M (ML-10M)~\footnote{https://grouplens.org/datasets/movielens/10m/} and Wikipedia~\footnote{https://github.com/clhchtcjj/BiNE/tree/master/data/wiki} are used in experiments. DBLP, ML-100K and ML-10M are adopted for top-K recommendation. Wikipedia is used for link prediction. We convert their user-item interaction matrices into the implicit data. The concrete statistics of them are listed in Table~\ref{dataset}. From it, we can observe that ML-10M is much larger than other datasets, since it is used to test whether our model can be deployed to large-scale bipartite graphs.
\subsubsection{Data Preprocessing}
As used in BiNE~\cite{Gao2018BiNEBN}, we select 60\% edges for training and remaining edges for test in both of DBLP and ML-10M. We use the same division in IGMC~\cite{Zhang2019IGMC} for ML-100K. Following experimental settings in the previous work~\cite{Gao2019LearningVR}, we split Wikipedia into two datasets, i.e., Wiki (5:5) and Wiki (4:6). The training/test ratios of these two datasets are 5:5 and 4:6, respectively.

\begin{table}
\centering
\caption{Statistics of datasets.}
\resizebox{8.0cm}{!}{
\begin{tabular}{|l|c|c|c|c|}
\hline
Datasets  & $|U|$ & $|V|$ & $|E|$ & Density  \\ \hline
DBLP  &6,001  &1,308 &29,256 &0.4\%  \\ \hline
ML-100K &943 &1,682 &100,000 &6.3\%  \\ \hline
ML-10M &69,878  &10,677  &10,000,054 &1.3\%   \\ \hline
Wikipedia &15,000  &3,214  &64,095 &0.1\%   \\ \hline

\end{tabular}
}
\label{dataset}
\vspace{-3mm}
\end{table}

\begin{table*}
\centering
\caption{Performance (\%) comparison of top-K recommendation on DBLP.}
\begin{tabular}{ccccccccccc}
\toprule
Model &F1@10 &NDCG@3 &NDCG@5 &NDCG@10  &MAP@3 &MAP@5 &MAP@10   &MRR@3 &MRR@5 &MRR@10 \\ \midrule
DeepWalk  &6.93   &4.91    &6.60   &9.12   &3.37     &4.23         &5.29   &9.04     &10.44        &11.70   \\
LINE      &8.45   &16.31   &19.03  &20.32  &14.25    &15.56        &16.07  &22.58    &25.28        &26.08   \\
Node2vec  &7.66   &20.33   &22.09  &23.00  &17.76    &18.61        &18.90  &28.00    
&29.21 &29.85 \\
VGAE      &10.16   &15.71   &16.57  &18.75  &11.08     &11.38         &12.17   &16.93    &18.30        &19.64   \\
Metapath2vec     &8.16   &19.81    &21.89   &22.70   &17.24     &18.15         &18.46   &27.23     &29.25         &29.68   \\ 
DMGI      &9.16   &19.71   &22.01  &23.65  &17.09    &18.27        &18.87  &26.69    &28.27        &30.13   \\
PinSage   &$\underline{12.55}$     &18.62      &21.17     &23.97     &14.71       &16.04           &17.30     &27.75       &29.68           &30.84 \\
BiNE      &11.36  &19.85   &21.95  &25.15  &17.12    &18.05        &19.34  &27.14    &29.40        &31.33   \\
GC-MC     &12.02  &19.87   &22.18  &24.62  &16.75    &17.98        &19.12  &28.91    &30.65        &31.70   \\
IGMC      &12.18  &20.35   &22.65  &25.17  &17.21    &18.43        &$\underline{19.61}$  &29.56    &31.30        &32.36   \\
NeuMF     &11.14     &19.59      &21.08     &24.31     &16.46       &17.19           &18.44     &27.01       &28.23           &30.32 \\
\specialrule{0pt}{0pt}{0pt}  NGCF      &12.38     &$\underline{21.29}$      &$\underline{23.38}$     &$\underline{25.58}$     &$\underline{17.36}$       &$\underline{18.50}$   &19.51     &$\underline{30.96}$       &$\underline{32.48}$           &$\underline{33.44}$ \\ 
\specialrule{0pt}{2pt}{0pt} \midrule
\specialrule{0pt}{2pt}{0pt}  
BiGI     &$\bm{14.27}^{*}$     &$\bm{23.56}^{*}$      &$\bm{25.39}^{*}$     &$\bm{28.28}^{*}$   &$\bm{19.10}^{*}$  &$\bm{20.15}^{*}$  &$\bm{21.49}^{*}$  &$\bm{33.19}^{*}$  &$\bm{35.35}^{*}$ &$\bm{36.51}^{*}$ \\ \bottomrule
\end{tabular}
\begin{center}
* indicates that the improvements are statistically significant for p < 0.05 judged by paired t-test.
\end{center}
\label{dblp}
\end{table*}

\begin{table*}
\centering
\caption{Performance (\%) comparison of top-K recommendation on ML-100K.}

\begin{tabular}{ccccccccccc}
\toprule
Model &F1@10 &NDCG@3 &NDCG@5 &NDCG@10  &MAP@3 &MAP@5 &MAP@10   &MRR@3 &MRR@5 &MRR@10  \\ \midrule
DeepWalk      &14.20 &7.17    &9.32   &13.13  &2.72    &3.54        &4.92  &43.86    &46.83        &48.75   \\
LINE          &13.71 &6.52    &8.57   &12.37  &2.45    &3.26        &4.67  &44.16    &44.37        &46.30   \\   
Node2vec      &14.13  &7.69  &9.91  &13.41  &3.07  &3.90  &5.19  &44.80  &48.02  &49.78 \\
VGAE          &11.38 &6.43    &8.18   &10.93  &2.35    &2.95        &3.94  &39.39    &42.32        &43.68   \\
Metapath2vec         &14.11  &7.88    &9.87   &13.35   &2.85    &3.71        &5.08  &45.49     &48.74         &49.83   \\ 
DMGI          &19.58  &10.16    &13.13   &18.31   &3.98    &5.33        &7.82  &59.33     &61.37         &62.71   \\
PinSage     &$\underline{21.68}$     &10.95      &$\underline{14.51}$     &20.27     &$\underline{4.52}$  &$\underline{6.18}$  &$\underline{9.13}$     &$\underline{62.56}$       &$\underline{64.77}$           &$\underline{65.76}$ \\ 
BiNE          &14.83 &7.69    &9.96   &13.79  &2.87    &3.80        &5.24  &48.14    &50.94        &52.51   \\
GC-MC         &20.65 &10.88   &13.87  &19.21  &4.41    &5.84        &8.43  &60.60    &62.21        &63.53   \\
IGMC          &18.81 &9.21    &12.20  &17.27  &3.50    &4.82       &7.18  &56.89    &59.13        &60.46  \\
NeuMF     &17.03     &8.87      &11.38     &15.89     &3.46       &4.54           &6.45     &54.42       &56.39           &57.79 \\
\specialrule{0pt}{0pt}{0pt} NGCF     &21.64     &$\underline{11.03}$      &14.49     &$\underline{20.29}$     &4.49       &6.15           &9.11     &$\underline{62.56}$       &64.62           &65.55 \\ 
\specialrule{0pt}{2pt}{0pt} \midrule
\specialrule{0pt}{2pt}{0pt} 
BiGI      &$\bm{23.36}^{*}$    &$\bm{12.50}^{*}$  &$\bm{15.92}^{*}$  &$\bm{22.14}^{*}$    &$\bm{5.41}^{*}$        &$\bm{7.15}^{*}$  &$\bm{10.50}^{*}$    &$\bm{66.01}^{*}$  &$\bm{67.70}^{*}$  &$\bm{68.78}^{*}$ \\ \bottomrule
\end{tabular}
\begin{center}
* indicates that the improvements are statistically significant for p < 0.05 judged by paired t-test.
\end{center}
\label{ml100k}
\end{table*}

\begin{table*}
\centering
\caption{Performance (\%) comparison of top-K recommendation on ML-10M.}
\begin{tabular}{ccccccccccc}
\toprule
Model &F1@10 &NDCG@3 &NDCG@5 &NDCG@10  &MAP@3 &MAP@5 &MAP@10   &MRR@3 &MRR@5 &MRR@10  \\ \midrule
DeepWalk      &7.25 &3.12    &4.39   &6.50  &1.12    &1.65        &2.55  &19.14    &20.97        &22.45   \\
LINE          &6.93 &3.07    &4.21   &6.24  &1.09    &1.55        &2.37  &19.69    &21.54        &23.08   \\   
Node2vec      &6.36  &2.82  &3.84  &5.71  &1.00  &1.40  &2.14  &18.10  &19.83  &21.32 \\
VGAE          &11.82 &5.00    &6.97   &10.61  &1.88    &2.79        &4.65  &34.75    &37.13        &39.00   \\
Metapath2vec  &8.28  &3.26    &4.66   &7.21   &1.18    &1.79        &2.98  &19.99     &21.92         &23.50   \\
DMGI          &12.52  &6.03    &8.09   &11.69   &2.15    &3.04        &4.77  &42.78     &44.86         &46.08   \\
PinSage     &14.93     &$\underline{7.53}$      &$\underline{10.07}$     &$\underline{14.14}$     &$\underline{2.70}$  &3.81  &5.85    &45.72      &47.58          &48.96 \\ 
GC-MC         &14.74 &7.05   &9.42  &13.73  &2.58    &3.68        &5.88  &48.07    &49.95        &51.18   \\
IGMC          &13.68 &6.58    &8.70  &12.78 &2.41    &3.32       &5.22  &45.57   &47.82        &49.29  \\
NeuMF     &13.91     &6.58      &8.92     &12.93     &2.38       &3.41           &5.34     &45.82       &48.14           &49.57\\
\specialrule{0pt}{0pt}{0pt} NGCF     &$\underline{15.11}$     &7.21      &9.67     &14.01     &2.67       &$\underline{3.84}$           &$\underline{6.16}$     &$\underline{48.19}$       &$\underline{50.15}$           &$\underline{51.33}$ \\ 
\specialrule{0pt}{2pt}{0pt} \midrule
\specialrule{0pt}{2pt}{0pt} 
BiGI      &$\bm{16.12}^{*}$   &$\bm{7.96}^{*}$  &$\bm{10.41}^{*}$  &$\bm{15.25}^{*}$    &$\bm{3.02}^{*}$        &$\bm{4.31}^{*}$  &$\bm{6.77}^{*}$    &$\bm{49.86}^{*}$  &$\bm{50.66}^{*}$  &$\bm{51.70}^{*}$ \\ \bottomrule
\end{tabular}
\begin{center}
* indicates that the improvements are statistically significant for p < 0.05 judged by paired t-test.
\end{center}
\label{ml10m}
\end{table*}

\subsection{Experimental Setting}
\subsubsection{Evaluation Metrics} In top-K recommendation, for each user, we first filter out some items that the user has already interacted with in training process. Then, we rank remaining items and evaluate ranking results with the following evaluation metrics: $F1$ score, $NDCG$ (Normalized Discounted Cumulative Gain), $MAP$ (Mean Average Precision) and $MRR$ (Mean Reciprocal Rank). All of these metrics are widely used in recommendation tasks. Two common metrics are used to evaluate the results of link prediction: $AUC$-$ROC$ (area under the ROC curve) and $AUC$-$PR$ (area under the Precison-Recall curve).
\subsubsection{Compared Baselines} We compare our model with the following strong baselines which can be divided into: 
\begin{itemize}
\item \textbf{Homogeneous graph embedding}: DeepWalk~\cite{Perozzi2014DeepWalkOL}, LINE \\~\cite{Tang2015LINELI}, Node2vec~\cite{Grover2016node2vecSF} and VGAE~\cite{kipf2016variational}. DeepWalk and Node2vec are typically random-walk based. 
LINE learns a joint probability distribution of connected nodes, and LINE (2nd) is exploited here due to its expressive performances. Based on variational auto-encoder~\cite{kingma2013auto}, VGAE adopts the graph convolutional network (GCN)~\cite{kipf2016semi} as the basic encoder to learn graph-structured data. 
\item \textbf{Heterogeneous graph embedding}: Metapath2vec~\cite{dong2017metapath2vec} and DMGI~\cite{Park2019UnsupervisedAM}. Metapath2vec first designs the metapath-based random walks to construct heterogeneous node neighborhoods. It then leverages a heterogeneous skip-gram model to learn node embeddings. DMGI~\cite{Park2019UnsupervisedAM} also follows the principle of MI maximization, and it uses the same infomax objective in DGI~\cite{Velickovic2018DeepGI}. 
\item \textbf{Bipartite graph embedding}: PinSage~\cite{ying2018graph} and BiNE~\cite{Gao2018BiNEBN}. PinSage integrates random walk into GNN architectures for high-scalable performances. BiNE jointly optimizes explicit and implicit relations in a unified framework. 
\item \textbf{Matrix completion}: GC-MC~\cite{berg2017graph} and IGMC~\cite{Zhang2019IGMC}. GC-MC introduces a relation-aware graph auto-encoder to learn embeddings of users and items. These representations are then used to reconstruct the rating links through a bilinear decoder. IGMC proposes a novel GNN based on local subgraphs for the task of inductive matrix completion. 
\item \textbf{Collaborative filtering}: NeuMF~\cite{He2017NeuralCF} and NGCF~\cite{NGCF19}. NeuMF uses MLP to learn the nonlinear interactions between user and item embeddings. NGCF considers the high-order connectivity via the proposed embedding propagation layer. 
\end{itemize}
\begin{table}
\centering
\caption{Performance comparison (\%) of link prediction.}
  \begin{tabular}{cccccc}
    \toprule
    \multirow{2}{*}{Model} & \multicolumn{2}{c}{Wiki (5:5)} & \multicolumn{2}{c}{Wiki (4:6)}  \\ \cmidrule(lr){2-3}\cmidrule(l){4-5}
    &AUC-ROC &AUC-PR &AUC-ROC &AUC-PR \\  \midrule
    DeepWalk & 87.19 & 85.30 & 81.60 & 80.29  \\
    LINE & 66.69 & 71.49 & 64.28 & 69.89  \\
    Node2vec &89.37 &88.12 &88.41 &87.55 \\
    VGAE & 87.81 & 86.93 & 86.32 & 85.74  \\
    Metapath2vec & 87.20 & 84.94 & 86.75 & 84.63  \\
    DMGI  & 93.02 & 93.11 & 92.01 & 92.14 \\
    PinSage  & 94.27 & 93.95 & 92.79 & 92.56 \\
    BiNE  &$\underline{94.33}$ & 93.93 & $\underline{93.15}$ & 93.34 \\ 
    GC-MC & 91.90 & 92.19 & 91.40 & 91.74  \\
    IGMC & 92.85 & 93.10 & 91.90 & 92.19  \\
    NeuMF  & 92.62 & 93.38 & 91.47 & 92.63 \\
    \specialrule{0pt}{0pt}{0pt} NGCF  & 94.26 &$\underline{94.07}$ & 93.06 & $\underline{93.37}$ \\ \specialrule{0pt}{2pt}{0pt} \midrule
    BiGI   &$\bm{94.91}^{*}$ &$\bm{94.75}^{*}$ &$\bm{94.08}^{*}$ &$\bm{94.02}^{*}$ \\ \bottomrule
    \end{tabular}
\begin{center}
* indicates that the improvements are statistically significant for p < 0.05 judged by paired t-test.
\end{center}
\vspace{-3mm}
\label{wiki}
\end{table}
\subsubsection{Implementation Details}
PinSage is implemented by ourselves. Except from it, we use official implementations of other methods. To make a fair comparison, the side information of nodes is not exploited in all experiments. The embedding size $d$ is fixed as 128, the learning rate is 0.001 and all models are iterated with 100 epochs for convergence. For making a good trade-off between effectiveness and efficiency, we use 1-hop enclosing subgraphs as suggested by IGMC~\cite{Zhang2019IGMC}. The depth of our encoder $k$ (the number of stacked layers) is 2. The margin $\gamma$ used in Eq.(\ref{gnnloss}) is 0.3. the corruption rate $\beta$ is selected from \{1e-6, 1e-5, 1e-4, 1e-3, 1e-2, 1e-1\}, and the harmonic factor $\lambda$ is selected from 0.1 to 0.9 with step length 0.2.
\par
To verify whether the results of our model are statistically significant, we perform paired t-test for each dataset. In addition, the results of BiNE on ML-10M are not provided. Although we use the official implementation of BiNE~\footnote{https://github.com/clhchtcjj/BiNE} for large-scale bipartite graphs, the concrete results of it are hard to be well reproduced (The generation of node sequences has not finished within 72 hours).

\begin{table}
\centering
\caption{Performance (\%) comparison of model variants.}
\begin{tabular}{ccccc}
\toprule
Model &F1@10 &NDCG@10  &MAP@10  &MRR@10 \\ \midrule
Encoder                                         &12.29  &25.71  &19.31  &31.08  \\
BiGI (node)                                &12.64  &25.46  &19.34  &34.08  \\ 
BiGI (pair)                                &13.23  &27.57  &21.20  &34.80  \\
BiGI (w/o att)                             &13.26  &27.59  &21.26  &35.33  \\
\specialrule{0pt}{0pt}{0pt} BiGI           &14.27  &28.28  &21.49  &36.51  \\ 
\specialrule{0pt}{2pt}{0pt} \midrule
\specialrule{0pt}{2pt}{0pt} VGAE            &10.16  &18.75  &12.17  &19.64  \\ 
\specialrule{0pt}{0pt}{0pt} BiGI (VGAE)    &11.15  &24.36  &19.05  &30.68  \\ 
\specialrule{0pt}{2pt}{0pt} \midrule
\specialrule{0pt}{2pt}{0pt} NGCF            &12.38  &25.58  &19.51  &33.44  \\
BiGI (NGCF)                                &13.03  &26.25  &20.34  &35.23  \\
\specialrule{0pt}{2pt}{0pt} \midrule
\specialrule{0pt}{2pt}{0pt} PinSage            &12.55  &23.97  &17.30  &30.84  \\
BiGI (PinSage)                                &13.26  &27.61  &21.48  &35.11  \\\bottomrule
\end{tabular}
\label{ablation}
\end{table}

\subsection{Top-K Recommendation}
Table~\ref{dblp}, Table~\ref{ml100k} and Table~\ref{ml10m} demonstrate the performances of compared methods on DBLP, ML-100K and ML-10M. The best performance is in boldface and the second is underlined. From them, we have the following observations. 1) Our method consistently yields the best performances on these datasets for all metrics. It demonstrates the high effectiveness of learning the global properties of bipartite graph. 2) Modeling the structural characteristics of bipartite graph is very important. Homogeneous and heterogeneous graph embeddings ignore such characteristics, and they are inferior to BiGI and to other bipartite graph embeddings. 
3) It should be noticed that DMGI also maximizes MI between local and global representations, but the performance of it is not satisfying. Therefore, designing a suitable infomax objective for bipartite graphs plays a central role in our work.

\begin{figure}
\begin{center}
\includegraphics[width=8.5cm,height=4cm]{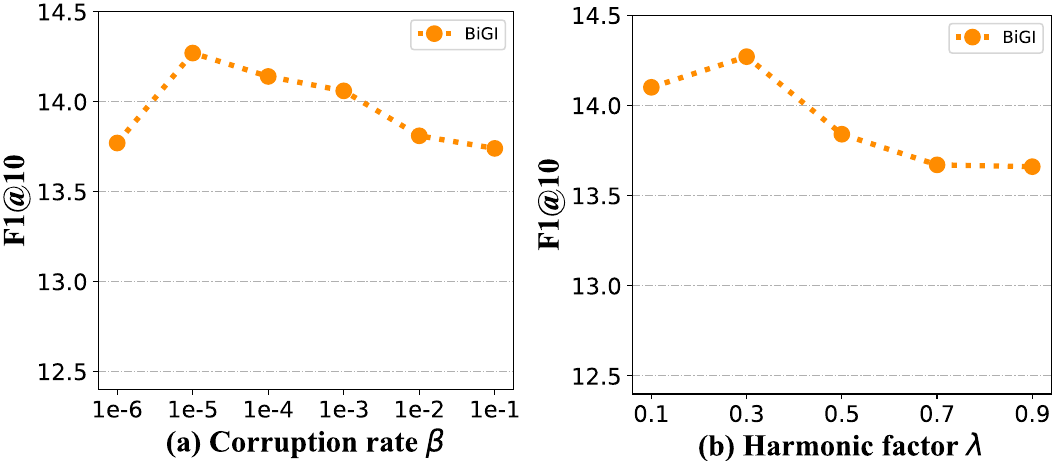}
\caption{Results of parameter sensitivity.}
\label{parameters}
\end{center}
\vspace{-3mm}
\end{figure}

\subsection{Link Prediction}
For the task of link prediction, given a node pair $(u_{i}, v_{j})$, we feed the corresponding embeddings $\bm{u}_{i}$ and $\bm{v}_{j}$ into a logistic regression classifier which is trained on the observed edges of bipartite graph. Table~\ref{wiki} shows the performances of all models, and our method achieves higher predictive results on both datasets. It demonstrates that the global properties of bipartite graph are beneficial to learn node representations. In particular, capturing long-range dependencies of heterogeneous nodes is helpful to down-stream tasks.

\subsection{Discussions of Model Variants}
We investigate the effects of different local representations and the extensibility of proposed infomax objective. The results of these model variants and an ablation study of the proposed encoder are provided in Table~\ref{ablation}. The experiments are conducted on DBLP. BiGI (node) uses each node embedding as the local representation. BiGI (pair) simply concatenates the representations of node pair $(u,v)$ as the local representation. BiGI (w/o att) calculates the representation of subgraph via the mean operation instead of the attention mechanism. BiGI (VGAE), BiGI (NGCF) and BiGI (PinSage) adopt VGAE, NGCF and PinSage as their encoders, respectively. All of them keep the same infomax objective with BiGI.
\par
From the results in Table~\ref{ablation}, we can draw the following conclusions. 1) The proposed encoder achieves competitive performance. Furthermore, in contrast with it, the improvements of BiGI are also significant. 2) Through the comparison of different local representations, we find that constructing a suitable local representation is crucial to BiGI. Introducing the subgraph-level attention mechanism into the calculation of local representation is a sensible choice. 3) By contrast with VGAE, NGCF and PinSage, the improvements of BiGI (VGAE), BiGI (NGCF) and BiGI (PinSage) are satisfying. It indicates that the proposed infomax objective can be seamlessly incorporated into other encoders to capture the global properties of bipartite graph. 

\subsection{Parameter Sensitivity} 
We investigate the parameter sensitivity of our model on DBLP with respect to two hyper-parameters: the corruption rate $\beta$ in Eq.(\ref{corruption}) and the harmonic factor $\lambda$ in Eq.(\ref{totalloss}). As shown in Figure~\ref{parameters}, when $\beta$=1e-5 and $\lambda=0.3$, our model achieves the best result. Therefore, choosing relative small values of $\beta$ and $\lambda$ is a reasonable way. Moreover, our model is robust to the changes of $\beta$ and $\lambda$. Even in the worst settings of $\beta$ and $\lambda$, BiGI is still better than other baselines shown in Table~\ref{dblp}.

\begin{figure}
\begin{center}
\includegraphics[width=8.5cm,height=4.4cm]{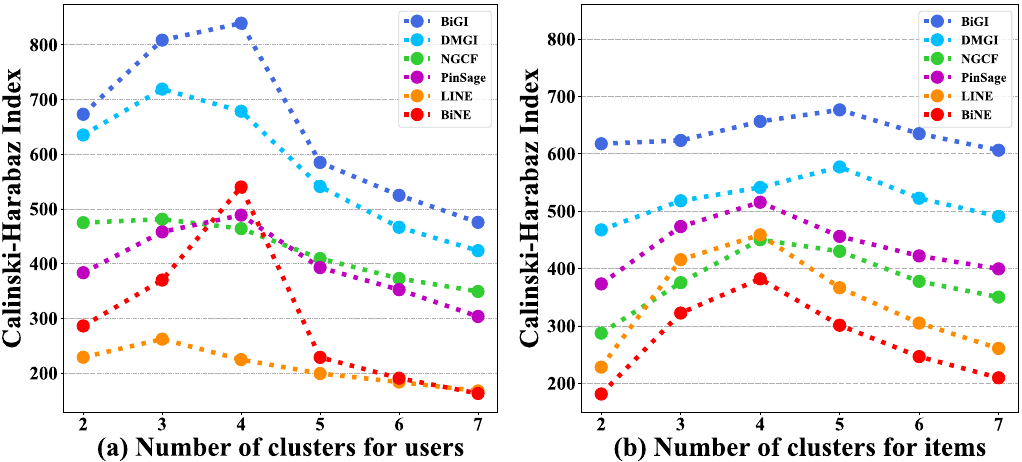}
\caption{Results of clustering analysis. BiGI achieves the best clustering results (A higher score is preferred).}
\label{cluster}
\end{center}
\vspace{-3mm}
\end{figure}

\subsection{Analysis of the Global Properties} 
In this section, to validate that our method is better to capture the global properties of bipartite graph, we conduct two detailed comparisons between BiGI and other strong baselines. In the first experiment, we provide two clustering analyses of users and items which are conducted on ML-100K. We first save all representations of users and items and then cluster them via the well-known K-Means algorithm. The clustering metric Calinski-Harabasz Index (CHI)~\cite{calinski1974dendrite} is used here. CHI measures the ratio between the within-cluster dispersion and the between-cluster dispersion. It is also commonly used to evaluate the task of community detection~\cite{liu2019evaluation,chowdhary2017community}.
As shown in Figure~\ref{cluster}, compared with other graph embeddings, BiGI achieves the best clustering results with the varying number of clusters. It demonstrates that BiGI can better capture community structures of users and items simultaneously. 
\par
\begin{figure}
\begin{center}
\includegraphics[width=8.5cm,height=4.4cm]{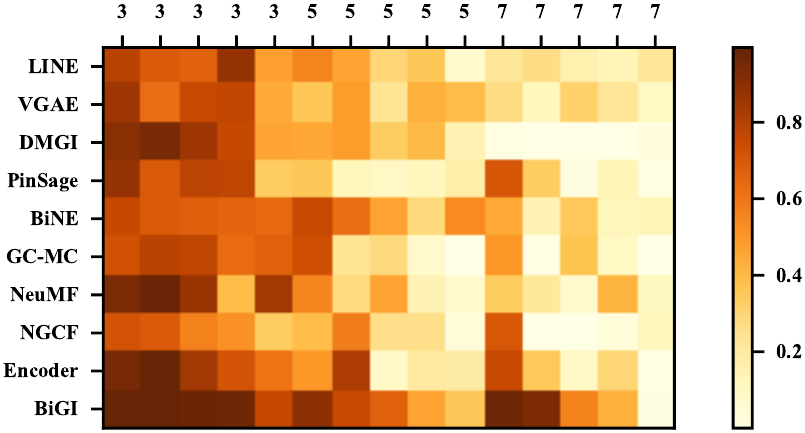}
\caption{Visualization of prediction scores. BiGI achieves the better results compared with other baselines.}
\label{heatmap}
\end{center}
\vspace{-3mm}
\end{figure}

Another comparison is used to test whether the long-range dependencies of heterogeneous nodes can be learned by our model. The scores are predicted by BiGI and several baselines for fifteen node pairs $\{(u_i, v_j)\}$ which are randomly picked from the test data of DBLP. These node pairs can be actually divided into three groups in terms of the distance between $u_i$ and $v_j$, i.e., 3, 5 and 7. From Figure~\ref{heatmap}, we have the following conclusions. 1) When the distance of target node pair is relative short, e.g. 3, all baselines and BiGI are capable of learning the latent interaction of node pair. 2) With the increase of distance, the observable relation between $u_i$ and $v_j$ is gradually weakened. Compared with state-of-the-art baselines, BiGI still maintains promising results. It demonstrates that BiGI can learn the long-range dependency of $u_i$ and $v_j$ even though they are distant from each other. 

\section{Conclusion}
In this paper, we propose a novel bipartite graph embedding named as BiGI. We first introduce a novel bipartite graph encoder to learn initial node representations. Two prototype representations are then generated via aggregating different homogeneous node information, which are further used to construct the global representation. Furthermore, we incorporate the structure prior into local representations via the designed subgraph-level attention mechanism. Through maximizing the MI between local and global representations, BiGI can recognize the global properties of bipartite graph effectively. Extensive experiments demonstrate that BiGI consistently outperforms state-of-the-art baselines on various datasets for different tasks. 

\section*{Acknowledgement}
This research was supported by the National Key Research and Development Program of China (grant No.2016YFB0801003), the Strategic Priority Research Program of Chinese Academy of Sciences (grant No.XDC02040400) and the National Social Science Foundation of China (grant No.19BSH022). Shu Guo and Tingwen Liu are corresponding authors.

\balance
\bibliographystyle{ACM-Reference-Format}
\bibliography{ref.bib}
\end{document}